\documentclass{aa}  
\usepackage[authoryear]{natbib}
\usepackage{graphicx}
\usepackage{amsmath}
\usepackage{amssymb}

\usepackage{txfonts}
\usepackage{url}
\begin{document}

\title{INTEGRAL observations of Her X-1
%\thanks{Based on observations with INTEGRAL, an ESA project with instruments and science data centre funded by ESA member states (especially the PI countries: Denmark, France, Germany, Italy, Switzerland, Spain), Czech Republic and Poland, and with the participation of Russia and the USA.}
}

\author{
D.~Klochkov\inst{1} \and 
R.~Staubert\inst{1} \and 
K.~Postnov\inst{3} \and
N.~Shakura\inst{3} \and 
A.~Santangelo\inst{1} \and
S.~Tsygankov\inst{2} \and 
A.~Lutovinov\inst{2} \and 
I.~Kreykenbohm\inst{1,4} \and 
J.~Wilms\inst{5}
}

\institute{
Institut f\"ur Astronomie und Astrophysik, University of T\"ubingen, Sand 1, 72076 T\"ubingen, Germany 
\and
Space Research Institute, Profsoyuznaya str. 84/32, 117997 Moscow, Russia
\and
Sternberg Astronomical Institute, Moscow University, 119992 Moscow, Russia
\and
INTEGRAL Science Data Centre, Chemin d'Ecogia, 16, 1290, Versoix, Switzerland
\and
Dr. Karl Remeis-Sternwarte, Astronomisches Institut, Universit\"at Erlangen-N\"urnberg, Sternwartstr. 7, 96049 Bamberg, Germany
}

\date{Accepted 29/01/2008}

\abstract
{}
{
We investigate the X-ray spectral and timing properties of the 
accreting X-ray pulsar \hbox{Her~X-1} observed with the \textsl{INTEGRAL} 
satellite in July--August 2005.
}
{
The data analyzed in this work cover a substantial part of one 
main-on state of the source. The short-time scale pulse period 
development is measured. X-ray pulse profiles for different energy 
ranges and time intervals are constructed. Pulse-averaged and 
pulse-phase resolved broad band X-ray spectra are studied. Spectral 
changes during X-ray dips are explored.
}
{
The X-ray pulse profiles are found to change significantly during 
the period of observations. 
For the first time a strong spin-up is measured within one 35~d cycle.
Spectral characteristics observed during the X-ray dips are consistent 
with their interpretaion as due to partial covering as has been reported 
by several authors. The fundamental cyclotron absorption line is 
firmly observed in both pulse-averaged and pulse-phase resolved 
X-ray spectra. The energy, width, and the depth of the line 
are found to vary significantly with pulse phase.
}
{}


\maketitle

%%-----------------------------------------------------------
\section{Introduction}

Discovered in 1972 by the {\em Uhuru} satellite 
\citep{Giacconi_etal73,Tananbaum_etal72} 
\hbox{Her~X-1} is one of the most intensively studied accreting pulsars. 
Being part of a low mass X-ray binary it shows strong variability 
on very different time scales: the 1.24\,s spin period of the neutron star, 
the 1.7\,d binary period, the 35\,d period of precession of the 
warped and tilted accretion disk \citep{GerendBoynton76,Shakura_etal99,
Ketsaris_etal00}, and the 1.65\,d period of the pre-eclipse dips 
\citep{Giacconi_etal73,Tananbaum_etal72}. 
Due to the high orbital inclination of the system ($i>80^{\circ}$) the 
counter-orbitally precessing warped accretion disk around the neutron 
star covers the X-ray source from the observer during a substantial 
part of the 35\,d period. This gives rise to the alternation of 
so-called {\em on} (high X-ray flux) and {\em off} (low X-ray flux) states. 
The 35\,d cycle contains two {\em on} states -- the {\em main-on} and 
the {\em short-on} -- separated by $\sim7-8$\,d {\em off} state. 
The X-ray flux in the middle of the main-on state is $\sim4-5$ 
times higher than that in the middle of the short-on. The sharp 
transition from the off-state to the main-on is called the {\em turn-on} 
of the source. Turn-ons are usually used for counting the cycles.  
The 35\,d period manifests itself also in variations of the shape of 
X-ray pulse profiles \citep{Soong_etal90, Truemper_etal86, Deeter_etal98, 
Scott_etal00} and modulation of optical light curves \citep{GerendBoynton76, 
HowarthWilson83}.

In spite of a large amount of available observational data, 
the physical interpretation of several observed phenomena in 
\hbox{Her~X-1} are still unclear. For example,
the physical mechanisms responsible for the disk 
precession, the X-ray dips, and the evolution of X-ray pulse profiles are 
highly debated. In this work we present X-ray observations of \hbox{Her~X-1} 
performed by the {\sl INTEGRAL} observatory. 

In Sect.~\ref{obs} we describe the observations and data processing.
In Sect.~\ref{herintpp} we present energy- and time-dependent 
pulse profiles. In Sect.~\ref{herintper} we explore the behavior
of the pulse period. The spectral analysis is presented in 
Sects.~\ref{heravspe} (pulse-averaged spectra), \ref{herintdips}
(spectral changes during X-ray dips), and \ref{prs} (pulse-resolved
analysis). The results are discussed in Sect.~\ref{discussion}.
In Sect.~\ref{concl} we present a summary and conclusions.

%%-------------------------------------------------------------
\section{Observations and data processing\label{obs}}

Her~X-1 was observed by {\sl INTEGRAL} \citep{Winkler_etal03}
on July 22 -- August 3, 2005 (MJD: 53573--53585
\footnote{MJD = JD -- 2400000.5}). The observations 
were spread over $\sim$5 orbital revolutions of the satellite: revs. 338, 
339, 340, 341, and 342. In our analysis we used data obtained with 
the instruments {\sl JEM-X} \citep{Lund_etal03}, {\sl IBIS/ISGRI} 
\citep{Lebrun_etal03,Ubertini_etal03}, 
and {\sl SPI} \citep{Vedrenne_etal03}.
We used the energy range $\sim$3--20~keV for \textsl{JEM-X} and
$\sim$20--100~keV for \textsl{ISGRI} and \textsl{SPI}
(the spectrum of \hbox{Her~X-1} is falling off steeply with energy
so that there is no signal above 100~keV).
The data processing was performed with the version 6.0 Offline 
Science Analysis (OSA) software distributed by ISDC 
\citep{Courvoisier_etal03}. An additional gain correction based 
on the analysis of the position of the tungsten background line 
was added (similar to \citealt{Tsygankov_etal06}). For the pulse-phase 
resolved spectroscopy we also used the software developed at IASF, 
Palermo for \textsl{IBIS/ISGRI}\footnote{\url{http://www.ifc.inaf.it/~ferrigno/INTEGRALsoftware.html}} 
\citep{Mineo_etal06,Ferrigno_etal07}. This software allows to construct 
phase-energy matrices from which pulse-resolved spectra and 
energy-resolved pulse profiles can easily be extracted.  
The results of the analysis of pulse phase averaged spectra were independently 
checked using the software package developed at Space Research Institute, 
Moscow.

\begin{figure}
\centering
\includegraphics[width=9cm]{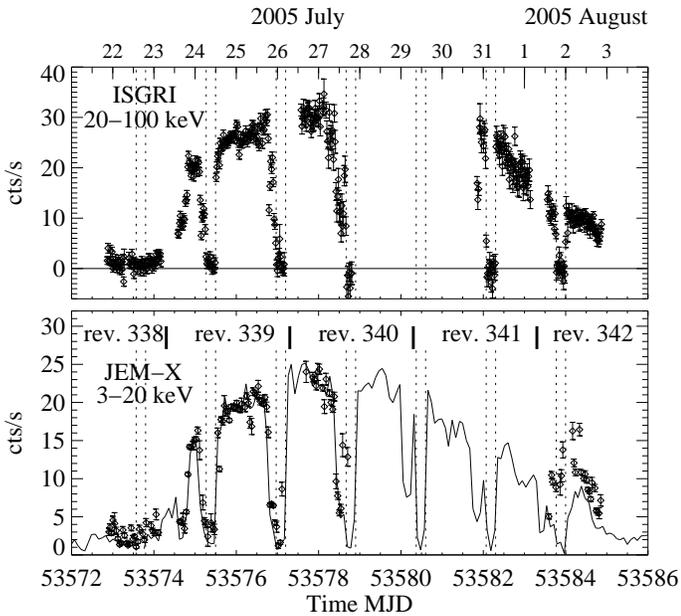}
\caption{X-ray light curves of \hbox{Her~X-1} obtained with {\sl ISGRI} 
(top panel) and {\sl JEM-X} (bottom panel). The solid curve in the 
bottom panel shows the {\sl ASM RXTE} light curve averaged over 
many 35\,d cycles \citep{Klochkov_etal06}. The {\sl INTEGRAL} revolution 
numbers are indicated.}
\label{lc}
\end{figure}

Figure~\ref{lc} shows the light curves of \hbox{Her~X-1} obtained 
with {\sl ISGRI} 
(top) and {\sl JEM-X} (bottom). The observations partially cover a 
main-on state of the source, 35\,d phases $\phi_{\rm pre} \sim$0--0.11 
and $\phi_{\rm pre} \sim$0.20--0.28 ($\phi_{\rm pre}=0$ is the phase of 
the turn-on of the source). The turn-on occurred at MJD $\sim 53574.7$, 
corresponding to orbital phase $\phi_{\rm orb} \sim 0.7$. 
The solid curve on top of the {\sl JEM-X} light curve is the {\sl ASM RXTE} 
\citep{Levine_etal96} light curve averaged 
over many 35\,d cycles and renormolized to match the one from 
{\sl JEM-X}. All {\sl JEM-X} data from revolution 341 are rejected 
by the ISDC team
because of a strong solar flare. It is seen that the light curve is 
quite typical for \hbox{Her~X-1}. Pre-eclipse dips (decrease of the flux just 
before the eclipse) are clearly seen.

%%---------------------------------------------------------------
\section{Timing analysis}

\begin{figure*}
  \centering
  \includegraphics[width=17cm]{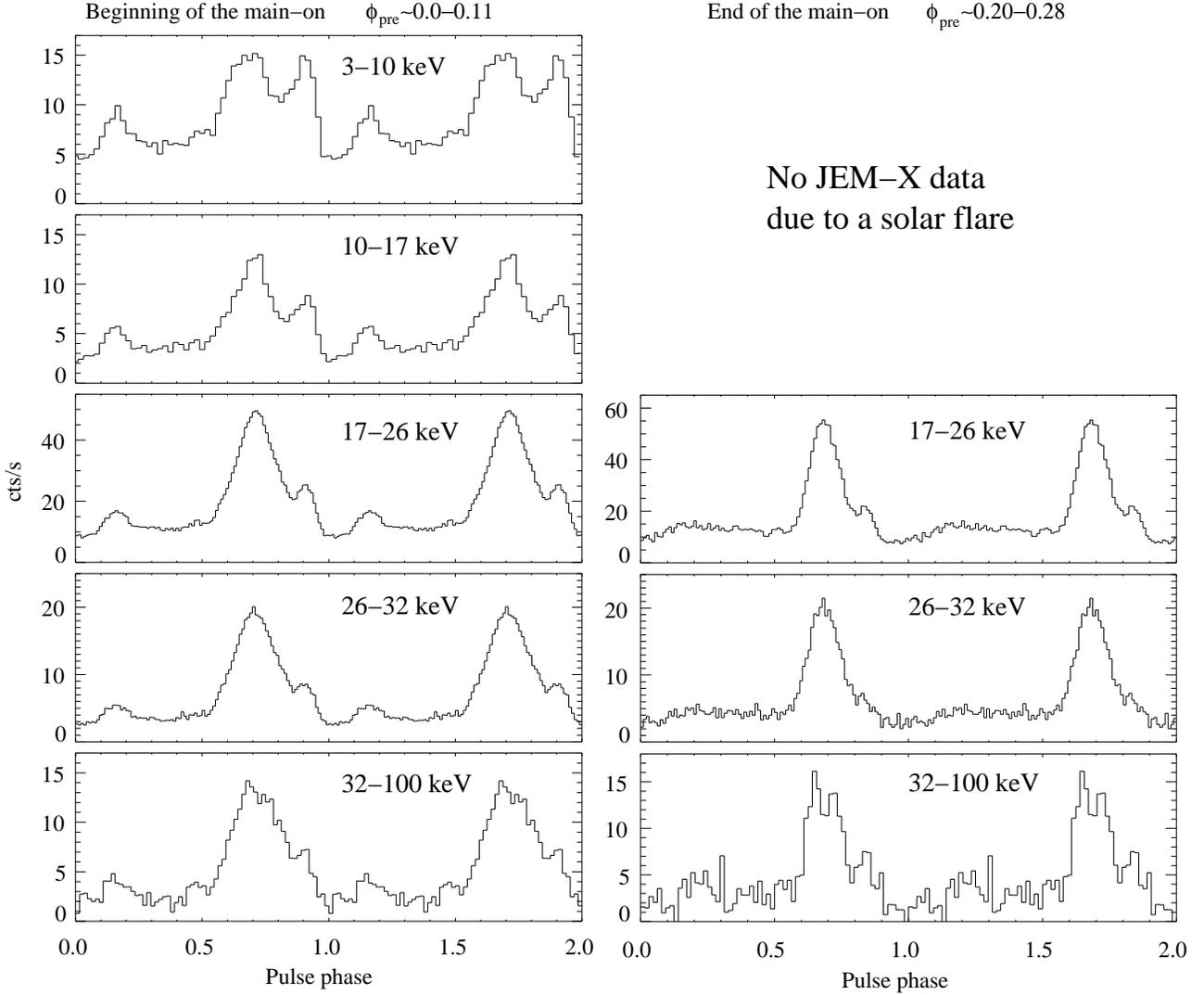}
  \caption{Energy-resolved X-ray pulse profiles of \hbox{Her~X-1} at the 
    start of the main-on (left column) and at the end of the 
    main-on (right column).}
  \label{ppdiss}
\end{figure*}

%%- - - - - - - - - - - - - - - - - - - - - - - - - - - - - - - -
\subsection{Pulse profiles\label{herintpp}}
To construct the 1.24\,s pulse profiles of the source all photon
arrival times were converted to the solar system barycenter and 
corrected for binary motion. The orbital parameters (Table~\ref{orbpar}) 
used for the binary correction are based on an updated ephemeris arrived 
at by combining historical timing data with the timing results of the
latest {\sl RXTE} observations of \hbox{Her~X-1} 
(the corresponding paper is being prepared). 
As a folding period for constructing pulse profiles we used 
$P_{\rm spin}=1.23775836(10)$~s which was obtained with the standard 
epoch folding method \citep{Leahy_etal83}.

It was found that the pulse profiles do not change significantly 
during the start (revs. 339 and 340) and during the end (rev. 341) 
of the main-on state. Therefore, we constructed the pulse profiles 
separately for these two intervals without further splitting in time. 
We excluded the data falling inside X-ray eclipses, pre-eclipse dips 
or located outside the main-on state (during the off-state the data 
are very noisy and show only marginal sine-like pulsations).
The resulting energy-resolved pulse profiles are shown in 
Fig.~\ref{ppdiss}. It is seen that the profiles change significantly 
from the start to the end of the main-on state. The main peak 
(pulse phase $\sim$0.6--0.8) is stronger and narrower at the end 
of the main-on. The interpulse (pulse phase $\sim$0.1--0.2)
almost disappears. This behavior is typical for \hbox{Her~X-1}: 
analogous pulse profiles were observed by \textsl{RXTE} at different 
times \citep[see e.g.][]{Kuster_etal05}. We also studied the energy
dependence of the pulse fraction $F$ determined as
\begin{equation}
F = \frac{ I_{\rm max} - I_{\rm min} }{I_{\rm max} + I_{\rm min}},
\end{equation}
where $I_{\rm max}$ and $I_{\rm min}$ are background-corrected count 
rates in the maximum and minimum of the pulse profile respectively. 
It is clearly seen (Fig.~\ref{frac}) that the pulse fraction increases 
with energy (there may be a saturation at energies above 
$\sim$30~keV, however the statistics is marginal).
The structure of the profile is also energy-dependent, 
going from a triple-peaked shape at lower energies to an almost 
single-peaked at higher energies. The main peak becomes narrower 
toward higher energies.
A deeper analysis of the pulse profiles including the modeling
of their shape (similar to that performed in \citealt{PanchenkoPostnov94}
but taking into account 35\,d variation) is ongoing and will be presented 
elsewhere.

%Table------------------------------------------------------------
\begin{table}
\caption{Orbital parameters of \hbox{Her~X-1} used to correct the arrival
times of photons for orbital motion in the binary.}
\label{orbpar}
\centering
\begin{tabular}{l l}
$P_{\rm orb}$     &= 1.700167233\,d,         \\
$T_\frac{\pi}{2}$ &= 53571.982111 (MJD),     \\
$a\sin i$         &= $13.1831\,{\rm light\,sec}$.  \\
\end{tabular}
\end{table}
%-----------------------------------------------------------------

%%- - - - - - - - - - - - - - - - - - - - - - - - - - - - - - - -
\subsection{Pulse period behavior\label{herintper}}

To explore more precisely the intrinsic (not affected by the orbital 
motion in the binary) spin period of the neutron star $P_{\rm spin}$ 
during the \textsl{INTEGRAL} observations we performed a
phase-connection analysis similar to 
\citet{Ferrigno_etal07, Deeter_etal81} using well defined average pulse 
profiles.
The length of the time intervals used to produce a single pulse profile was 
3--5 INTEGRAL Science Windows (6--10 ksec). This allowed to produce pulse 
profiles of a quality allowing to measure their shift in pulse phase with 
respect to other pulse profiles within $\sim$0.01~s uncertainties.
Figure~\ref{timing_int} shows the 
observed pulse arrival times minus the calculated times (using a constant 
pulse period). If we assign the same statistical uncertainty 
of 0.01s (determined from the scattering of data points) to each data point 
then the linear fit to the data corresponding to a constant pulse period 
(dashed line) gives $\chi^2_{\rm red}=1.9$ for 23 d.o.f. while the 
quadratic fit corresponding to the presence of a non-zero 
$\dot P_{\rm spin}$  results in $\chi^2_{\rm red}=1.0$ for 22 d.o.f.
We conclude, therefore, that the neutron star is spinning up with 
$-\dot P_{\rm spin} = (5.8\pm 1.5)\times 10^{-13} {\rm s/s}$. This value is
significantly higher than the averaged spin-up trend of \hbox{Her~X-1} 
which is $\sim 1.1\times 10^{-13}$~s/s. Around the average
spin-up strong variability of the pulse period 
from one main-on to the next
is known to exist \citep[see e.g. Fig.~1 in][]{Staubert_etal06a}.

%%---------------------------------------------------------------
\section{Spectral analysis}

The spectral analysis has been performed using the 
\texttt{XSPEC} v.11.3.2l spectral fitting package \citep{Arnaud96}. 
For the start of the main-on the data from all three X-ray instruments
have been used. The fit, however, is mainly driven by \textsl{JEM-X} 
($\lesssim 20$~keV) and \textsl{ISGRI} ($\gtrsim 20$~keV) data. 
The spectra obtained with \textsl{SPI} ($\gtrsim 20$~keV) have much lower 
statistics. For the end of the main-on only the higher energy part of the 
spectrum is available since the {\sl JEM-X} data could not be used 
due to a solar flare (see above). Following the OSA User 
Manuals\footnote{\url{http://isdc.unige.ch/index.cgi?Support+documents}}, 
we added systematic errors at a level of 2\% in quadrature 
to all \textsl{JEM-X} and \textsl{ISGRI} spectral points in order 
to account for uncertainties in the response matrices of 
the respective instruments.

%%- - - - - - - - - - - - - - - - - - - - - - - - - - - - - - - -
\subsection{Pulse-averaged spectra\label{heravspe}}

For constructing the pulse-averaged spectrum, we used the data from 
revolutions 339 and 340 (the start of the main-on state, see 
Fig.~\ref{lc}). During this interval the data from all three X-ray 
instruments are available. Eclipses and dips,
which are clearly seen in the light curve (Fig.~\ref{lc})
and have sharp boundaries, were excluded
from the analysis. To model the broad band spectral continuum
we used a power law with the exponential cutoff model \texttt{highecut} 
of \texttt{XSPEC}:
\begin{equation}
I_{\rm cont} \propto
\begin{cases}
E^{-\Gamma}, & \text{if\,} E \leq E_{\rm cutoff} \\
E^{-\Gamma} \cdot \exp{\left(-\frac{E-E_{\rm cutoff}}{E_{\rm fold}}\right)}, & \text{if\,}
E > E_{\rm cutoff},
\end{cases}
\label{highecut}
\end{equation}
where $E$ is the photon energy; $\Gamma,\,E_{\rm cutoff}$, and 
$E_{\rm fold}$ are model parameters. A Gaussian line is added to model 
the iron fluorescence line at $\sim$6.5~keV. The cyclotron absorption
feature is modeled by a multiplicative absorption line with a Gaussian 
optical depth profile. So, the final spectral function $I$ is the following:
\begin{equation*}
I=I_{\rm cont}\cdot
\exp\left\{-\tau_{\rm cycl}
\exp\left(-\frac{(E-E_{\rm cycl})^2}{2\sigma_{\rm cycl}^2}\right)\right\}+
\end{equation*}
\begin{equation}
\qquad + K\exp\left(-\frac{(E-E_{\rm Fe})^2}{2\sigma_{\rm Fe}^2}\right),
\end{equation}
where $E_{\rm cycl}$ and $\sigma_{\rm cycl}$ are the centroid 
energy and width of the cyclotron line, $E_{\rm Fe}$ and $\sigma_{\rm Fe}$ 
are the energy and width of the iron emission line. 
$\tau_{\rm cycl}$ and $K$ are the numerical constants describing the 
strength of the cyclotron line and iron line, respectively. The count rate 
spectrum is shown in Fig.~\ref{herspe}. The middle panel shows the 
residuals after fitting the spectrum with the continuum model 
without the iron emission line and the cyclotron absorption line. 
Systematic features around 6~keV and 40~keV are clearly seen.
The bottom panel shows the residuals of our final fit which
includes the two lines. 

\begin{figure}
  \resizebox{\hsize}{!}{\includegraphics{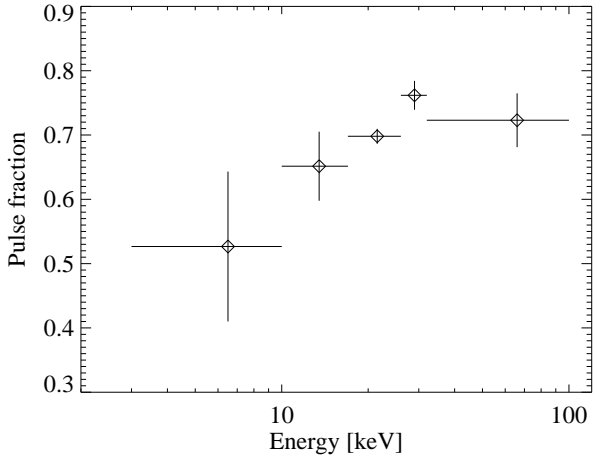}}
  \caption{Pulse fraction of \hbox{Her~X-1} as a function of energy.}
  \label{frac}
\end{figure}

\begin{figure}
  \resizebox{\hsize}{!}{\includegraphics{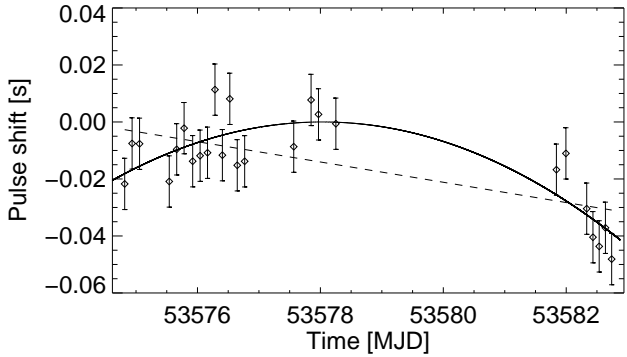}}
  \caption{Observed pulse arrival times of \hbox{Her~X-1} minus calculated 
    times using a constant pulse period. The dashed line shows a linear fit 
    corresponding to a constant pulse period while the solid line shows a 
    quadratic fit corresponding to a 
    $\dot P_{\rm spin} = (-5.8\pm 1.5)\times 10^{-13} {\rm s/s}$ .}
  \label{timing_int}
\end{figure}

%Table---------------------------------------------------------------------
\begin{table}
\caption{Best fit spectral parameters of \hbox{Her~X-1} 
for the observations of revs. 339 and 340.
1$\sigma$(68\%)-uncertainties ($\chi^2_{\rm min}+1$) for one parameter 
of interest are shown.}
\label{herfit}
\centering
\renewcommand{\arraystretch}{1.2}
\setlength{\tabcolsep}{1cm}
\vspace{8pt}
\begin{tabular}{l l}
\hline\hline
    Parameter             &   Value                \\
\hline
                          &                        \\
$\Gamma$                  & $0.91\pm 0.01$         \\
$E_{\rm cutoff}$ [keV]    & $25.5_{-0.3}^{+0.2}$   \\
$E_{\rm fold}$ [keV]      & $9.0_{-0.2}^{+0.3}$    \\
$E_{\rm Fe}$ [keV]        & $6.57_{-0.13}^{+0.11}$ \\
$\sigma_{\rm Fe}$ [keV]   & $0.50_{-0.16}^{+0.17}$ \\
$E_{\rm cycl}$            & $38.2_{-0.6}^{+0.8}$ \\
$\sigma_{\rm cycl}$       & $9.3_{-0.5}^{+0.4}$    \\
$\tau_{\rm cycl}$         & $0.63_{-0.06}^{+0.04}$ \\
$F_\text{JEM-X}$          & 1.0 (fixed)            \\
$F_\text{ISGRI}$          & $0.81_{-0.02}^{+0.01}$ \\
$F_\text{SPI}$            & $0.96\pm 0.02$         \\ 
$\chi^2_\text{red}$/d.o.f.& 1.19/207               \\
\hline
\end{tabular}
\end{table}
%------------------------------------------------------------------------

To account for large systematic uncertainties in 
the normalization of
the instruments we introduced in our models a free 
multiplicative factor for each instrument: $F_{\rm ISGRI}$, $F_{\rm SPI}$ 
and  $F_{\rm JEM-X}$ (for \textsl{JEM-X} the factor was fixed to 1.0). 
The best-fit parameters with corresponding 1$\sigma$(68\%)-uncertainties 
are listed in Table~\ref{herfit}. The position of the cyclotron line 
at $38.2_{-0.6}^{+0.8}$~keV is consistent with that measured by 
\textsl{RXTE} at a similar time (see Table~1 in \citealt{Staubert_etal07}) 
and, therefore, supports the correlation between the cyclotron line 
energy and the maximum main-on flux found by \citet{Staubert_etal07}. 

We have checked the presence and the centroid energy of the cyclotron line
using different continuum models (such as e.g. Fermi-Dirac
cutoff, \citealt{Tanaka86}) as well as using different line profiles
(e.g. Lorenzian profile). It was found that the presence
and the energy of the feature are independent of the choice of the 
continuum and the line profile model.
The \texttt{highecut} model was found to provide a better description of the
spectral continuum in \hbox{Her~X-1} with respect to other
models. Additionally, it allows to compare our results with previous
observations of the source, most of which were analyzed using the
\texttt{highecut} model \citep[see e.g.][]{Gruber_etal01}. A discontinuity
of the first derivative of the spectral function at 
$E_{\rm cutoff}\sim 25$~keV 
(which can be noticed in the residuals in Fig.~\ref{herspe})
is far from the cyclotron line energy and does not affect
the line parameters \citep[see however][]{Kretschmar_etal97}.

We also explored the pulse-averaged spectrum obtained during 
revolution 341 (end of the main-on). The continuum spectral parameters 
are poorly constrained 
in this case due to the absence of the low-energy ($\lesssim 20$\,keV) 
part of the spectrum. However, the found position of the cyclotron line 
$E_{\rm cycl} = 37.3_{-1.1}^{+1.3}$~keV is consistent with that measured in 
revolutions 339 and 340.

\begin{figure}
  \resizebox{\hsize}{!}{\includegraphics[angle=-90]{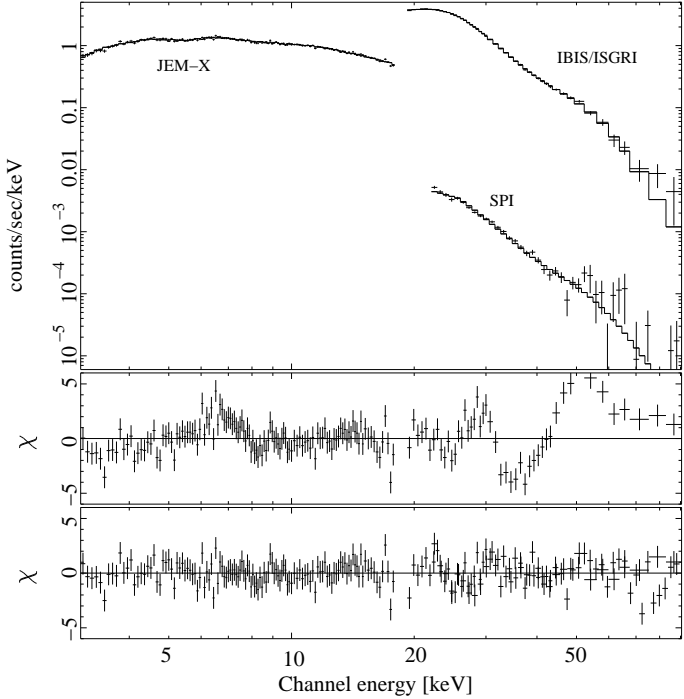}}
  \caption{{\em Top:} The pulse-averaged X-ray spectrum of \hbox{Her~X-1} obtained 
    during revolutions 339 and 340 (start of the main-on); {\em middle:} 
    residuals after fitting the spectrum with the continuum model described 
    by Eq.~\ref{highecut} without 
    %introducing 
    the iron emission line and 
    cyclotron absorption line (to avoid confusion the \textsl{SPI} data are 
    not shown in this panel, their statistics is lower than that of 
    \textsl{IBIS/ISGRI} and \textsl{JEM-X}); {\em bottom:} the 
    residuals after including the iron and cyclotron lines in the model.}
  \label{herspe}
\end{figure}

%%- - - - - - - - - - - - - - - - - - - - - - - - - - - - - - - -
\subsection{X-ray dips\label{herintdips}}

\begin{figure}
  \resizebox{\hsize}{!}{\includegraphics{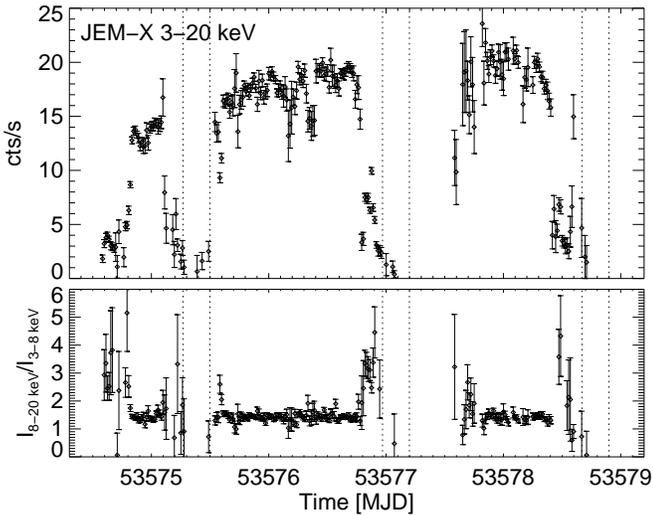}}
  \caption{The \textsl{JEM-X} 3--20\,keV light curve {\em (top panel)} 
    and the ratio of count rates in the harder (8--20~keV) and 
    softer (3--8~keV) energy ranges as a function of time 
    {\em (bottom panel)}.}
  \label{ratio}
\end{figure}

In the X-ray light curve taken with \textsl{INTEGRAL} one can see 
pre-eclipse dips on three subsequent orbits after the turn-on (Fig.~\ref{lc}). 
The turn-on of the source can be considered as egress from an
anomalous dip at $\phi_{\rm orb}\sim 0.5$ (according to 
the model of \citet{Shakura_etal99} the 
source is partially screened by the outer accretion disk rim in both 
cases: during the turn-on of the source and during anomalous dips).
Thus, the light curve obtained with 
\textsl{INTEGRAL} contains three pre-eclipse dips and one anomalous dip. 
Using the \textsl{JEM-X} data we calculated the ratio 
$H=I_{\rm 8-20\,keV}/I_{\rm 3-8\,keV}$, where $I_{\rm 8-20\,keV}$ and
$I_{\rm 3-8\,keV}$ are the count rates in the respective energy ranges.
The ratio $H$ as a function of time is shown in the bottom panel of 
Fig.~\ref{ratio}. It can be seen that $H$ increases during
X-ray dips most probably indicating the low-energy absorption by the cold
matter in the accretion stream (during the pre-eclipse dips) and in the outer
rim of the disk (during the anomalous dip). 
Even though the result is dominated by the last two pre-eclipse dips, the 
data of the first pre-eclipse dip and the anomalous dip are consistent.

\begin{figure}
  \resizebox{\hsize}{!}{\includegraphics[angle=-90]{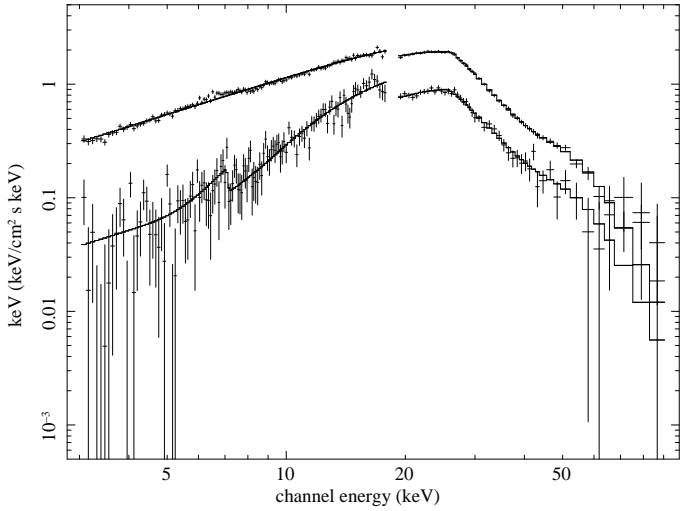}}
  \caption{Examples of pulse-averaged spectra of \hbox{Her~X-1} from the dips 
    (lower curve) and outside the dips (upper curve). The difference of the
    spectral shape at lower energies caused by photo-absorption inside 
    the dips is clearly seen. The step-like feature at $\sim$7~keV in the
    spectrum of the dips corresponds to the iron K-edge 
    (as implemented in the \texttt{phabs} model of \texttt{XSPEC}).}
  \label{dipspe}
\end{figure}

To check the absorption hypothesis we constructed \textsl{JEM-X} and 
\textsl{ISGRI} spectra corresponding to the four dips and compared them with 
those obtained outside the dips. 
We are aware of the fact that due to different formation 
mechanisms of anomalous and pre-eclipse dips their spectral
characteristics might be different. But due to the low statistics we
cannot perform a separate analysis of the two kinds of dips 
(the spectral parameters are not restricted in this case). 
To fit the spectrum from the dips we used 
the approach of \citet{Kuster_etal05}. It was assumed that a combination 
of direct and absorbed radiation is observed during the dips. Thus, we used 
a partial covering model which combines both, absorbed and non-absorbed 
spectra to fit the data during the dips. We used the same continuum model 
as in the previous section. The final spectral model can be written as
\begin{equation}
I(E) = A\cdot[1 + \alpha(E)\cdot I_{\rm cont}],
\label{parcovmod}
\end{equation}
where $I_{\rm cont}$ is the continuum model described in the previous 
section, and 
\begin{equation}
\alpha(E) = B\cdot e^{-N_{\rm H}\sigma_{\rm bf}(E)},
\end{equation}
where $\sigma_{\rm bf}(E)$ is the photoabsorption cross-section 
per hydrogen atom for matter of cosmic abundances 
\citep{Balucinska92}
used in the
\texttt{phabs} model of \texttt{XSPEC} and $N_{\rm H}$ is the equivalent 
hydrogen column density. A larger value of $B$ implies a larger degree of 
absorbed flux. From the fit of the spectrum of the dips we found 
$B=0.24\pm 0.03$ and $N_{\rm H}=(111_{-19}^{+18})\times 10^{22}~{\rm cm}^{-2}$. 
The best-fit parameters of the spectral continuum 
$I_{\rm cont}$ are consistent with those found from fitting the spectrum 
outside the dips (Table~\ref{herfit}). To compare this result with 
the spectrum outside the dips, the latter was also fit by the partial 
covering model (Eq.~\ref{parcovmod}). Both spectra (inside and outside 
the dips) are shown in Fig.~\ref{dipspe}. For the spectrum outside 
the dips (the upper one) the $N_{\rm H}$ parameter was found to be 
consistent with zero and the $B$ factor, consequently, was not restricted. 
We conclude therefore, that the X-ray data of \hbox{Her~X-1} during the dips 
are consistent with the partial absorption model.

%%- - - - - - - - - - - - - - - - - - - - - - - - - - - - - - - -
\subsection{Pulse-phase-resolved spectra\label{prs}}

\begin{figure}
  \resizebox{\hsize}{!}{\includegraphics[angle=-90]{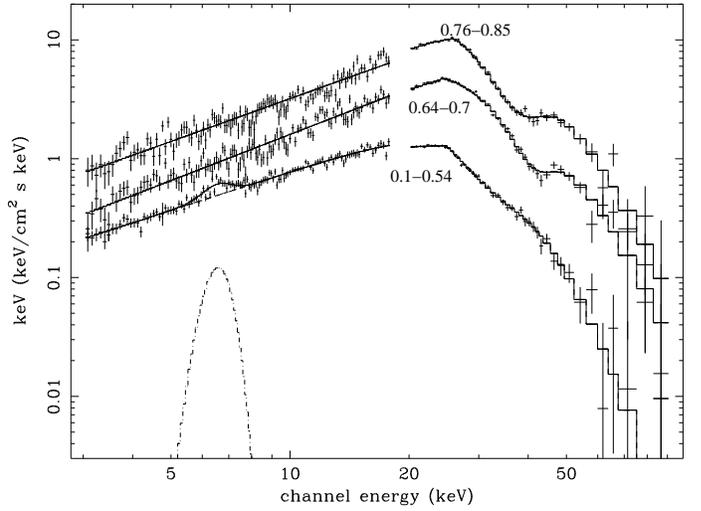}}
  \caption{Examples of pulse-resolved X-ray spectra of \hbox{Her~X-1} during 
    revolutions 339 and 340. Pulse phase intervals are indicated 
    (the pulse phase zero is the same as in Fig.~\ref{ppdiss}).}
  \label{dissprs3}
\end{figure}

It is well known that the X-ray spectrum of \hbox{Her~X-1} varies with 1.24\,s 
pulse phase \citep{Voges_etal82,Voges84,Soong_etal90b}. So, we have 
performed a separate analysis of the spectra accumulated in different 
pulse phase intervals. Figure~\ref{dissprs3} shows an example of pulse 
resolved spectra of the source. Variability of the continuum and the 
cyclotron line is clearly seen. 

Since the shape of the pulse profile is changing significantly from the 
start of the main-on (revs. 339, 340) to its end (rev. 341), 
we analyzed the data from these two intervals separately. 
Phase binning in each case was chosen to provide a similar statistics 
in each spectral bin. Pulse phase zero is the same as in 
Fig.~\ref{ppdiss}. The X-ray spectrum of each phase bin was 
fitted with the spectral model described in Sect.~\ref{heravspe}.
Figures~\ref{prs_cont} and \ref{prs_crsf} show the best-fit spectral 
parameters as a function of pulse phase for revs. 339 and 340 
{\em(left)} and the rev. 341 {\em(right)}. Vertical error bars correspond to 
1$\sigma$(68\%)-uncertainties. The dotted line shows the corresponding pulse 
profile. The spectral parameters in different pulse phase bins are also listed
in Table~\ref{prstable}.
Since no \textsl{JEM-X} data are available for rev. 341
the power law photon index $\Gamma$ could not be restricted by the fit in 
this case. The iron line at $\sim$6.5~keV was detected only in the phase bin 
0.1--0.54, corresponding to the off-pulse interval. The data from other phase 
intervals do not require inclusion of the line in the spectral model
to obtain a good fit. For the data corresponding to revs. 339 and 340,
it is seen that all continuum and the cyclotron line parameters are highly
variable with pulse phase. For revolution 341 the picture is not so
clear due to the absence of the low-energy ($\lesssim 20$~keV) part of the 
spectrum. Also the total exposure time and the average flux of the source in 
rev. 341 are smaller than those during revolutions 339 and 340.

%Table------------------------------------------------------------
\begin{table*}
\caption{Best fit spectral parameters of \hbox{Her~X-1} in different
pulse phase intervals. 1$\sigma$(68\%)-uncertainties 
($\chi^2_{\rm min}+1$) for one parameter of interest are shown.}
\label{prstable}
\centering
\renewcommand{\arraystretch}{1.2}
\begin{tabular}{l l l l l l l}
Phase bin &$\Gamma$              &$E_{\rm cutoff}$    & $E_{\rm fold}$        &$E_{\rm cycl}$      & $\sigma_{\rm cycl}$  & $\tau_{\rm cycl}$    \\
\hline\hline
Revs. 339 and 340    &           &                    &                       &                    &                      &                      \\
0.10--0.54&$0.92_{-0.04}^{+0.02}$&$25.8_{-0.9}^{+0.7}$&$ 6.35_{-0.20}^{+0.43}$&$29.6_{-0.5}^{+0.9}$&$6.86_{-0.31}^{+0.93}$&$0.67_{-0.11}^{+0.11}$\\ 
0.54--0.64&$1.06_{-0.02}^{+0.02}$&$24.3_{-0.3}^{+0.5}$&$10.11_{-0.38}^{+0.45}$&$39.6_{-0.9}^{+1.1}$&$4.32_{-1.09}^{+1.64}$&$0.42_{-0.06}^{+0.07}$\\ 
0.64--0.70&$0.73_{-0.04}^{+0.02}$&$24.3_{-0.2}^{+0.2}$&$ 9.59_{-0.31}^{+0.30}$&$40.7_{-0.5}^{+0.6}$&$4.84_{-0.62}^{+0.72}$&$0.66_{-0.06}^{+0.06}$\\ 
0.70--0.76&$0.48_{-0.02}^{+0.04}$&$23.2_{-0.3}^{+0.2}$&$ 9.48_{-0.24}^{+0.25}$&$40.8_{-0.4}^{+0.4}$&$5.69_{-0.48}^{+0.45}$&$0.98_{-0.06}^{+0.06}$\\ 
0.76--0.85&$0.79_{-0.02}^{+0.04}$&$25.2_{-0.6}^{+0.8}$&$ 9.90_{-0.29}^{+0.28}$&$37.7_{-0.5}^{+0.4}$&$5.30_{-0.69}^{+0.74}$&$0.71_{-0.05}^{+0.06}$\\ 
0.85--0.95&$0.97_{-0.09}^{+0.04}$&$26.2_{-0.6}^{+0.6}$&$ 7.86_{-0.23}^{+0.47}$&$32.9_{-1.4}^{+1.3}$&$9.63_{-1.01}^{+1.48}$&$0.71_{-0.10}^{+0.14}$\\ 
\hline
Rev. 341  &                      &                    &                       &                    &                      &                      \\
0.00--0.56&           ---        &$22.1_{-0.4}^{+0.3}$&$ 8.23_{-0.11}^{+0.18}$&$33.5_{-0.6}^{+0.1}$&$0.18_{-0.17}^{+0.60}$&$\geq 1.0$            \\
0.56--0.64&           ---        &$27.4_{-0.6}^{+0.8}$&$11.45_{-0.60}^{+0.58}$&$42.7_{-0.8}^{+0.7}$&$9.52_{-0.65}^{+1.14}$&$1.39_{-0.13}^{+0.13}$\\
0.64--0.72&           ---        &$25.2_{-0.5}^{+0.2}$&$ 9.68_{-0.37}^{+0.35}$&$39.5_{-0.3}^{+0.3}$&$5.67_{-0.23}^{+0.25}$&$0.92_{-0.07}^{+0.06}$\\
0.72--0.79&           ---        &$26.8_{-0.6}^{+0.5}$&$ 9.54_{-0.42}^{+0.28}$&$36.9_{-0.2}^{+0.4}$&$5.82_{-0.44}^{+0.44}$&$1.04_{-0.06}^{+0.07}$\\
0.79--1.00&           ---        &$23.8_{-0.5}^{+0.4}$&$ 8.14_{-0.35}^{+0.30}$&        ---         &            ---       &          ---         \\
\hline                           
\end{tabular}
\end{table*} 
%-----------------------------------------------------------------

%%---------------------------------------------------------------
\section{Discussion\label{discussion}}

\subsection{Pulse profiles and pulse period\label{discusspp}}

In Sect.~\ref{herintpp} we constructed 1.24\,s X-ray pulse profiles 
of \hbox{Her~X-1}. The shape of the profiles is both energy- and 
time-dependent. At higher energies the main peak becomes narrower 
and the pulsed fraction increases (Figs.~\ref{ppdiss} and~\ref{frac}). 
Such a dependence of the profile on energy is typical for accreting 
pulsars \citep[see e.g.][]{Tsygankov_etal07}. It can be understood 
in a simple purely geometrical picture: if the rotation axis of the 
neutron star is inclined with respect to the axis of its magnetic field, 
it is expected that the upper part of the accretion column emitting 
softer photons is seen during a larger part of the neutron star spin 
period while the emission region of harder photons, the ``base'' 
of the accretion column, is screened by the neutron star surface 
during most of the spin period. Additionally, the emission diagram 
of harder photons most probably originating close to the base of
the accretion column is believed to be narrower than that of softer photons 
due to the strong dependence of the scattering cross-section 
on the angle between the direction of a photon and the magnetic field
lines \citep[see e.g.][]{BaskoSunyaev76}. 

The difference of the pulse profiles in the left and right columns of 
Fig.~\ref{ppdiss} (corresponding to the start and the end of the
main-on state, respectively) shows the evolution of the profiles with 
35\,d phase. Two models have been proposed to explain these variations. 
\citet{Scott_etal00} suggested that the change of the profile from the 
start to the end of the main-on state is due to a 
``resolved occultation'' of the emitting regions on the neutron star 
by the precessing accretion disk. According to this model, the disk 
progressively occults the neutron star towards the end of the main-on. 
The other model assumes free precession of the neutron
star \citep{Truemper_etal86,Kahabka87,Kahabka89,
Prokhorov_etal90,Shakura_etal98b}. 
In this model the observed behavior of the profile is explained 
assuming that the emission region on the star surface has a complex 
shape (due to the presence of higher multipole components of the
neutron star's magnetic field). Changing the viewing conditions of 
the emitting region with the phase of the free precession causes 
the observed variation of the pulse profile 
\citep{Ketsaris_etal00,Wilms_etal03}. In pulse profiles obtained 
with {\sl INTEGRAL} one can see that the count rate at the maximum of 
the main peak is higher at the end of the main-on than that during 
the start of the main-on (Fig.~\ref{ppdiss}). In the case of 
a progressive occultation one would only expect a decrease in the 
intensity of any feature in the profile towards the end of the main-on. 
In the model of free precession, however, viewing conditions of 
different parts of the emission region may change with the precessional 
phase of the star in such a way that the intensity of particular 
features of the profile (e.g of the main peak) 
will increase towards the end of the main-on. We argue, therefore, that 
the {\sl INTEGRAL} observations analyzed in this work 
%lend some 
support the model of a freely precessing neutron star as an explanation of 
the time variation of the pulse profile in \hbox{Her~X-1}. 

In Sect.~\ref{herintper} we determined the time derivative of the 
intrinsic (not affected by the orbital motion in the binary) pulse period 
of the neutron star in \hbox{Her~X-1} during the \textsl{INTEGRAL} observation. 
The pulsar was found to spin-up with the rate 
$-\dot P_{\rm spin} = (5.8\pm 1.5)\times 10^{-13}~{\rm s/s}$. 
This value is more than five times larger than the mean spin-up trend 
of \hbox{Her~X-1} ($\sim 1.1\times 10^{-13}$~s/s) and consistent with the 
variations of $P_{\rm spin}$ between subsequent 35~d cycles,
e.g. as seen by {\sl BATSE} \citep{Kunz96,Nagase89,Sunyaev_etal88}
and \textsl{RXTE} \citep{Staubert_etal06a}.
However, this is the first time that such strong pulse period variations
have been measured within one 35\,d cycle.
A deeper analysis of these pulse period variations and their corelation
with the X-ray luminosity (including the data from \textsl{INTEGRAL}
presented here) is currently being performed and will be presented in a 
separate paper.

\subsection{Absorption during X-ray dips}

In Sect.~\ref{herintdips} we have shown that the X-ray spectrum 
of \hbox{Her~X-1} obtained during X-ray dips can be modeled using a 
partial covering model which assumes that the observed 
spectrum is a combination of direct and absorbed radiation. Such a 
combination has already been observed in \hbox{Her~X-1} during X-ray 
turn-ons \citep{DavisonFabian77,Becker_etal77,Parmar_etal80,
Kuster_etal05} 
and also during X-ray dips \citep[][although the observations 
included only the lower energy part of the spectrum 
$\lesssim$20~keV]{VrtilekHalpern85}. 

According to the model of \citet{Shakura_etal99} and 
\citet{Klochkov_etal06} (which we adopt in this work), most of 
the X-ray dips are produced by the occultation of the X-ray source by 
the cold matter in the accretion stream which moves out of the 
system's orbital plane. If the stream is not homogeneous but consists 
of ``blobs'' of material then the source will be screened during
some time intervals within a dip. Between these intervals the 
source will be visible or seen through less dense material. This 
picture seems to be confirmed by the complicated behavior of the 
light curve inside the dips (Fig.~\ref{ratio}). In our analysis,
in order to construct an X-ray spectrum of acceptable quality
we had to accumulate the flux during complete dips. The spectrum 
in this case will contain a superposition of the absorbed and 
non-absorbed flux. The partial covering model fits the spectrum of 
the dips clearly better than the simple absorption model.

\subsection{Pulse-phase variability of the X-ray spectrum}

\begin{figure}
  \resizebox{\hsize}{!}{\includegraphics{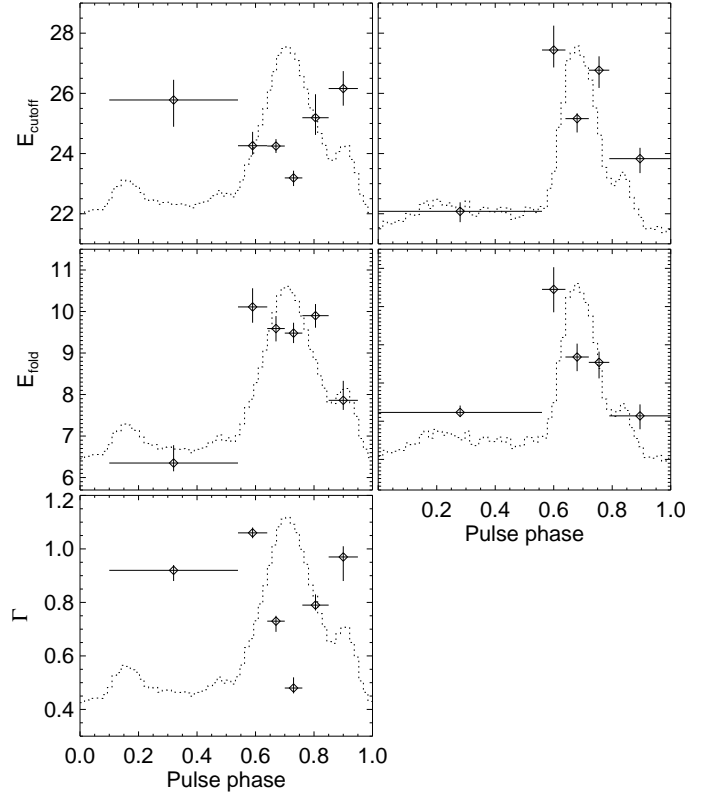}}
  \caption{Best-fit spectral continuum parameters as functions of the pulse 
    phase for revolutions 339 and 340 {\em(left)} and the revolution 341 
    {\em(right)}. The dashed curve shows the corresponding pulse profile. 
    Vertical error bars correspond to 1$\sigma$(68\%)-uncertainties.}
  \label{prs_cont}
\end{figure}

\begin{figure}
  \resizebox{\hsize}{!}{\includegraphics{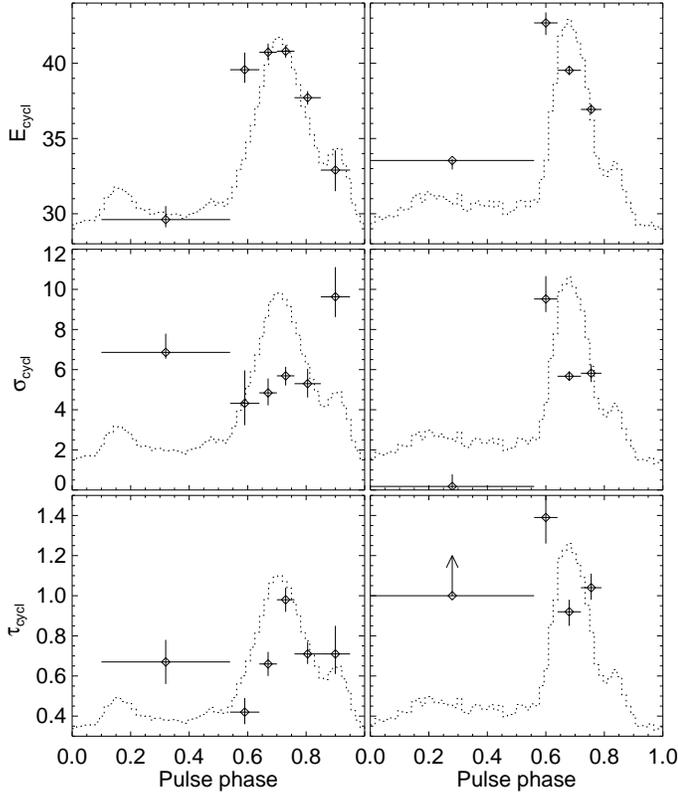}}
  \caption{The same as in Fig.~\ref{prs_cont} but for the cyclotron line 
    parameters (left: revolutions 339 and 340; right: revolution 341).}
  \label{prs_crsf}
\end{figure}

The dependence of the spectral parameters in \hbox{Her~X-1} on pulse phase
observed by \textsl{INTEGRAL} basically confirms the results obtained earlier
with \textsl{HEAO-1} \citep{Soong_etal90} but provide better restriction
of continuum parameters. 
The spectral variation with pulse phase is a common feature in X-ray 
pulsars \citep[see e.g.][and references therein]{Kreykenbohm_etal04}
which is usually attributed to the change of the viewing angle of 
the accretion region on the neutron star surface. Below we show that 
the pulse-phase variability observed in \hbox{Her~X-1} is basically consistent 
with this interpretation. 

As one can see in Fig.~\ref{prs_cont}, the power law photon index 
$\Gamma$ decreases during the main peak. It ranges from 
$\sim$1.0 in the off-pulse to $\sim$0.5 close to the maximum of the 
main peak \citep[see also][]{Lutovinov_etal00}. 
This effect reflects the sharpening of the main peak with energy 
which is observed in energy-resolved pulse profiles constructed 
in Sect.~\ref{herintpp} (Fig.~\ref{ppdiss}). The spectral hardening 
in the main peak can be explained by the dependence of the optical depth 
on the angle between the line-of-sight and the magnetic field lines 
\citep{Pravdo_etal77}. The closer the viewing direction is to the 
magnetic axis, the deeper we look into the emission region where harder 
photons originate. Since it is generally accepted that the main peak 
corresponds to the radiation from one of the two magnetic poles, the 
viewing direction is closest to the magnetic axis during the main peak
which causes the observed effect.

The exponential folding energy $E_{\rm fold}$ 
is much higher during the main peak than during the rest of the
pulsation cycle.
In comptonized X-ray spectra (which are normally observed 
from accreting pulsars) this parameter is believed to be proportional 
to the plasma temperature \citep[see e.g.][]{RybickiLightman79}. As 
already mentioned, during the main peak we see the emission from higher 
optical depth. This means that the observed radiation originates closer 
to the base of the accretion column where the plasma temperature 
is higher.

The cyclotron line centroid energy $E_{\rm cycl}$ is also 
higher during the
main peak of the profile (Fig.~\ref{prs_crsf}). A relative amplitude 
of its variation is $\sim$25\%. 
As argued from $\Gamma$ and $E_{\rm fold}$,
during the main peak we see X-rays that originate closer to the neutron
star surface, i.e. where the magnetic field strength is higher. The 
increase in $E_{\rm cycl}$ during the main peak is, therefore, 
qualitatively consistent with this picture.
If one assumes that the radiation comes from a compact accretion column 
located at the magnetic pole of the neutron star with a pure dipole field 
then the change in the height of the observed emission region necessary 
to produce a $\sim$25\% variation of the magnetic field strength 
is $\sim$1.1~km (for a neutron star radius of 10~km). 
However, as it was shown by \citet{Staubert_etal07b}, the height
of the emission and line forming region in case of 
\hbox{Her~X-1} is most probably much smaller, $\sim 10^4$~cm.
Thus, a changing height of the emitting region above the neutron star 
cannot explain the observed variability of the cyclotron line energy 
with pulse phase in this source.

Another possibility for the line energy variability is to assume a complicated 
structure of the magnetic field at the site of X-ray emission. 
As shown by \citet{Shakura_etal91} and \citet{PanchenkoPostnov94}, 
the complex shape of the observed pulse profiles in \hbox{Her~X-1}
suggests that such a complicated field structure including higher multipole 
components is indeed present in the source. In this case, 
at different rotational phases of the neutron star we will observe 
emission coming from the regions corresponding to different sub-structures 
of the non-dipole magnetic field with very different field strengths. 
Furthermore, the numerical modeling of \hbox{Her~X-1} pulse profiles 
observed with \textsl{EXOSAT} performed by 
\citet{PanchenkoPostnov94}
suggests that the magnetic field stength at the ring-like structure
on the neutron star's surface corresponding to the quadrupole 
component of the field (which was introduced in the model)
is indeed by $\sim$27\% lower than that at the magnetic pole.
Thus, a complicated magnetic field structure on the surface of the 
neutron star can cause the observed 25\%-variation of $E_{\rm cycl}$ 
with pulse phase.

%%---------------------------------------------------------------
\section{Summary and conclusion\label{concl}}

In this work we analyzed the {\sl INTEGRAL} observations of the 
accreting X-ray pulsar \hbox{Her~X-1}. X-ray pulse profiles of the
source are constructed and their energy- and time-dependence are
discussed. A strong spin-up is found during the {\sl INTEGRAL} observations. 
This is the first time that pulse period variation is measured within one
main-on state of the source.
The value of the spin-up rate is consistent with
the typical difference between the values of the pulse period
previously found in adjacent 35\,d cycles. 
Spectral changes during X-ray dips are studied. The X-ray spectrum of 
the dips was modeled by a partial covering model which assumes that 
the observed spectrum is a combination of direct and absorbed radiation. 
This is in agreement with the previously suggested interpretation
assuming that
the obscuring matter (accretion stream) is not
homogeneous but rather consists of blobs of material
\citep[see e.g.][]{VrtilekHalpern85}. Energy-resolved 
X-ray pulse profiles as well as the variation of cyclotron
line and continuum parameters with 1.24\,s pulse phase were explored.
The spectral changes with pulse phase are shown to be qualitatively 
(for the cyclotron line energy -- also quantitatively)
consistent with those expected from the viewing conditions of 
the complex emitting region which vary with the rotational phase of 
the neutron star.

\begin{acknowledgements}
This research is based on observations with INTEGRAL, an ESA project with instruments and science data centre funded by ESA member states (especially the PI countries: Denmark, France, Germany, Italy, Switzerland, Spain), Czech Republic and Poland, and with the participation of Russia and the USA.
The work was supported by the DFG grants Sta 173/31-2 and 
436 RUS 113/717/0-1 and the corresponding RBFR grants 
RFFI-NNIO-03-02-04003 and RFFI 06-02-16025, as well as 
DLR grant 50 0R 0302 and RFFI 07-02-01051.
We also thank ISSI (Bern, Switzerland) for its hospitality during the 
team meetings of our collaboration.
AL acknowledges the financial support of the 
Russian Science Support Foundation.
\end{acknowledgements}

\bibliographystyle{aa}
\bibliography{refs}

\begin{thebibliography}{55}
\expandafter\ifx\csname natexlab\endcsname\relax\def\natexlab#1{#1}\fi

\bibitem[{{Arnaud}(1996)}]{Arnaud96}
{Arnaud}, K.~A. 1996, in Astronomical Society of the Pacific Conference Series,
  Vol. 101, Astronomical Data Analysis Software and Systems V, ed. G.~H.
  {Jacoby} \& J.~{Barnes}, 17

\bibitem[{{Ba\l{}uci\'nska-Church} \& {McCammon}(1992)}]{Balucinska92}
{Ba\l{}uci\'nska-Church}, M. \& {McCammon}, D. 1992, ApJ, 400, 699

\bibitem[{{Basko} \& {Sunyaev}(1976)}]{BaskoSunyaev76}
{Basko}, M.~M. \& {Sunyaev}, R.~A. 1976, MNRAS, 175, 395

\bibitem[{{Becker} {et~al.}(1977){Becker}, {Boldt}, {Holt}, {Pravdo},
  {Rothschild}, {Serlemitsos}, {Smith}, \& {Swank}}]{Becker_etal77}
{Becker}, R.~H., {Boldt}, E.~A., {Holt}, S.~S., {et~al.} 1977, ApJ, 214, 879

\bibitem[{{Courvoisier} {et~al.}(2003){Courvoisier}, {Walter}, {Beckmann},
  {Dean}, {Dubath}, {Hudec}, {Kretschmar}, {Mereghetti}, {Montmerle},
  {Mowlavi}, {Paltani}, {Preite Martinez}, {Produit}, {Staubert}, {Strong},
  {Swings}, {Westergaard}, {White}, {Winkler}, \&
  {Zdziarski}}]{Courvoisier_etal03}
{Courvoisier}, T.~J.-L., {Walter}, R., {Beckmann}, V., {et~al.} 2003, A\&A,
  411, L53

\bibitem[{{Davison} \& {Fabian}(1977)}]{DavisonFabian77}
{Davison}, P.~J.~N. \& {Fabian}, A.~C. 1977, MNRAS, 178, 1P

\bibitem[{{Deeter} {et~al.}(1981){Deeter}, {Pravdo}, \&
  {Boynton}}]{Deeter_etal81}
{Deeter}, J.~E., {Pravdo}, S.~H., \& {Boynton}, P.~E. 1981, \apj, 247, 1003

\bibitem[{{Deeter} {et~al.}(1998){Deeter}, {Scott}, {Boynton}, {Miyamoto},
  {Kitamoto}, {Takahama}, \& {Nagase}}]{Deeter_etal98}
{Deeter}, J.~E., {Scott}, D.~M., {Boynton}, P.~E., {et~al.} 1998, ApJ, 502, 802

\bibitem[{{Ferrigno} {et~al.}(2007){Ferrigno}, {Segreto}, {Santangelo},
  {Wilms}, {Kreykenbohm}, {Denis}, \& {Staubert}}]{Ferrigno_etal07}
{Ferrigno}, C., {Segreto}, A., {Santangelo}, A., {et~al.} 2007, A\&A, 462, 995

\bibitem[{{Gerend} \& {Boynton}(1976)}]{GerendBoynton76}
{Gerend}, D. \& {Boynton}, P.~E. 1976, ApJ, 209, 562

\bibitem[{{Giacconi} {et~al.}(1973){Giacconi}, {Gursky}, {Kellogg}, {Levinson},
  {Schreier}, \& {Tananbaum}}]{Giacconi_etal73}
{Giacconi}, R., {Gursky}, H., {Kellogg}, E., {et~al.} 1973, ApJ, 184, 227

\bibitem[{{Gruber} {et~al.}(2001){Gruber}, {Heindl}, {Rothschild}, {Coburn},
  {Staubert}, {Kreykenbohm}, \& {Wilms}}]{Gruber_etal01}
{Gruber}, D.~E., {Heindl}, W.~A., {Rothschild}, R.~E., {et~al.} 2001, ApJ, 562,
  499

\bibitem[{{Howarth} \& {Wilson}(1983)}]{HowarthWilson83}
{Howarth}, I.~D. \& {Wilson}, B. 1983, MNRAS, 202, 347

\bibitem[{{Kahabka}(1987)}]{Kahabka87}
{Kahabka}, P. 1987, NASA STI/Recon Technical Report N, 88, 19405

\bibitem[{{Kahabka}(1989)}]{Kahabka89}
{Kahabka}, P. 1989, in ESA SP-296: Two Topics in X-Ray Astronomy, Volume 1: X
  Ray Binaries. Volume 2: AGN and the X Ray Background, ed. J.~{Hunt} \&
  B.~{Battrick}, 447--452

\bibitem[{Ketsaris {et~al.}(2000)Ketsaris, Kuster, Postnov,
  {et~al.}}]{Ketsaris_etal00}
Ketsaris, N.~A., Kuster, M., Postnov, K., {et~al.} 2000, in Proc. Int. Workshop
  "Hot Points in Astrophysics", JINR, Dubna, p. 192, ed. V.~Belyaev,
  arXiv:astro-ph/0010035

\bibitem[{{Klochkov} {et~al.}(2006){Klochkov}, {Shakura}, {Postnov},
  {Staubert}, {Wilms}, \& {Ketsaris}}]{Klochkov_etal06}
{Klochkov}, D.~K., {Shakura}, N.~I., {Postnov}, K.~A., {et~al.} 2006, Astronomy
  Letters, 32, 804

\bibitem[{{Kretschmar} {et~al.}(1997){Kretschmar}, {Kreykenbohm}, {Wilms},
  {Staubert}, {Maisack}, {Kendziorra}, {Heindl}, {Rothschild}, {Gruber}, \&
  {Grove}}]{Kretschmar_etal97}
{Kretschmar}, P., {Kreykenbohm}, I., {Wilms}, J., {et~al.} 1997, in ESA Special
  Publication, Vol. 382, The Transparent Universe, ed. C.~{Winkler}, T.~J.-L.
  {Courvoisier}, \& P.~{Durouchoux}, 141

\bibitem[{{Kreykenbohm} {et~al.}(2004){Kreykenbohm}, {Wilms}, {Coburn},
  {Kuster}, {Rothschild}, {Heindl}, {Kretschmar}, \&
  {Staubert}}]{Kreykenbohm_etal04}
{Kreykenbohm}, I., {Wilms}, J., {Coburn}, W., {et~al.} 2004, A\&A, 427, 975

\bibitem[{Kunz(1996)}]{Kunz96}
Kunz, M. 1996, PhD thesis, University of T\"ubingen, Germany

\bibitem[{{Kuster} {et~al.}(2005){Kuster}, {Wilms}, {Staubert}, {Heindl},
  {Rothschild}, {Shakura}, \& {Postnov}}]{Kuster_etal05}
{Kuster}, M., {Wilms}, J., {Staubert}, R., {et~al.} 2005, A\&A, 443, 753

\bibitem[{{Leahy} {et~al.}(1983){Leahy}, {Elsner}, \&
  {Weisskopf}}]{Leahy_etal83}
{Leahy}, D.~A., {Elsner}, R.~F., \& {Weisskopf}, M.~C. 1983, ApJ, 272, 256

\bibitem[{{Lebrun} {et~al.}(2003){Lebrun}, {Leray}, {Lavocat}, {Cr{\'e}tolle},
  {Arqu{\`e}s}, {Blondel}, {Bonnin}, {Bou{\`e}re}, {Cara}, {Chaleil}, {Daly},
  {Desages}, {Dzitko}, {Horeau}, {Laurent}, {Limousin}, {Mathy}, {Mauguen},
  {Meignier}, {Molini{\'e}}, {Poindron}, {Rouger}, {Sauvageon}, \&
  {Tourrette}}]{Lebrun_etal03}
{Lebrun}, F., {Leray}, J.~P., {Lavocat}, P., {et~al.} 2003, A\&A, 411, L141

\bibitem[{{Levine} {et~al.}(1996){Levine}, {Bradt}, {Cui}, {Jernigan},
  {Morgan}, {Remillard}, {Shirey}, \& {Smith}}]{Levine_etal96}
{Levine}, A.~M., {Bradt}, H., {Cui}, W., {et~al.} 1996, ApJ, 469, L33

\bibitem[{{Lund} {et~al.}(2003){Lund}, {Budtz-J{\o}rgensen}, {Westergaard},
  {Brandt}, {Rasmussen}, {Hornstrup}, {Oxborrow}, {Chenevez}, {Jensen},
  {Laursen}, {Andersen}, {Mogensen}, {Rasmussen}, {Om{\o}}, {Pedersen},
  {Polny}, {Andersson}, {Andersson}, {K{\"a}m{\"a}r{\"a}inen}, {Vilhu},
  {Huovelin}, {Maisala}, {Morawski}, {Juchnikowski}, {Costa}, {Feroci},
  {Rubini}, {Rapisarda}, {Morelli}, {Carassiti}, {Frontera}, {Pelliciari},
  {Loffredo}, {Mart{\'{\i}}nez N{\'u}{\~n}ez}, {Reglero}, {Velasco}, {Larsson},
  {Svensson}, {Zdziarski}, {Castro-Tirado}, {Attina}, {Goria}, {Giulianelli},
  {Cordero}, {Rezazad}, {Schmidt}, {Carli}, {Gomez}, {Jensen}, {Sarri},
  {Tiemon}, {Orr}, {Much}, {Kretschmar}, \& {Schnopper}}]{Lund_etal03}
{Lund}, N., {Budtz-J{\o}rgensen}, C., {Westergaard}, N.~J., {et~al.} 2003,
  A\&A, 411, L231

\bibitem[{{Lutovinov} {et~al.}(2000){Lutovinov}, {Grebenev}, {Pavlinsky}, \&
  {Sunyaev}}]{Lutovinov_etal00}
{Lutovinov}, A.~A., {Grebenev}, S.~A., {Pavlinsky}, M.~N., \& {Sunyaev}, R.~A.
  2000, Astronomy Letters, 26, 691

\bibitem[{{Mineo} {et~al.}(2006){Mineo}, {Ferrigno}, {Foschini}, {Segreto},
  {Cusumano}, {Malaguti}, {di Cocco}, \& {Labanti}}]{Mineo_etal06}
{Mineo}, T., {Ferrigno}, C., {Foschini}, L., {et~al.} 2006, A\&A, 450, 617

\bibitem[{{Nagase}(1989)}]{Nagase89}
{Nagase}, F. 1989, PASJ, 41, 1

\bibitem[{{Panchenko} \& {Postnov}(1994)}]{PanchenkoPostnov94}
{Panchenko}, I.~E. \& {Postnov}, K.~A. 1994, A\&A, 286, 497

\bibitem[{{Parmar} {et~al.}(1980){Parmar}, {Sanford}, \&
  {Fabian}}]{Parmar_etal80}
{Parmar}, A.~N., {Sanford}, P.~W., \& {Fabian}, A.~C. 1980, MNRAS, 192, 311

\bibitem[{{Pravdo} {et~al.}(1977){Pravdo}, {Boldt}, {Holt}, \&
  {Serlemitsos}}]{Pravdo_etal77}
{Pravdo}, S.~H., {Boldt}, E.~A., {Holt}, S.~S., \& {Serlemitsos}, P.~J. 1977,
  \apjl, 216, L23

\bibitem[{{Prokhorov} {et~al.}(1990){Prokhorov}, {Shakura}, \&
  {Postnov}}]{Prokhorov_etal90}
{Prokhorov}, M.~E., {Shakura}, N.~I., \& {Postnov}, K.~A. 1990,
  {Thirty-five-day cycle of HER X-1: Synthesised optical light curves in the
  model of freely processing neutron star}, Tech. rep.

\bibitem[{{Rybicki} \& {Lightman}(1979)}]{RybickiLightman79}
{Rybicki}, G.~B. \& {Lightman}, A.~P. 1979, {Radiative processes in
  astrophysics} (New York, Wiley-Interscience, 1979.~393 p.)

\bibitem[{{Scott} {et~al.}(2000){Scott}, {Leahy}, \& {Wilson}}]{Scott_etal00}
{Scott}, D.~M., {Leahy}, D.~A., \& {Wilson}, R.~B. 2000, ApJ, 539, 392

\bibitem[{{Shakura} {et~al.}(1991){Shakura}, {Postnov}, \&
  {Prokhorov}}]{Shakura_etal91}
{Shakura}, N.~I., {Postnov}, K.~A., \& {Prokhorov}, M.~E. 1991, Soviet
  Astronomy Letters, 17, 339

\bibitem[{{Shakura} {et~al.}(1998){Shakura}, {Postnov}, \&
  {Prokhorov}}]{Shakura_etal98b}
{Shakura}, N.~I., {Postnov}, K.~A., \& {Prokhorov}, M.~E. 1998, A\&A, 331, L37

\bibitem[{{Shakura} {et~al.}(1999){Shakura}, {Prokhorov}, {Postnov}, \&
  {Ketsaris}}]{Shakura_etal99}
{Shakura}, N.~I., {Prokhorov}, M.~E., {Postnov}, K.~A., \& {Ketsaris}, N.~A.
  1999, A\%A, 348, 917

\bibitem[{{Soong} {et~al.}(1990{\natexlab{a}}){Soong}, {Gruber}, {Peterson}, \&
  {Rothschild}}]{Soong_etal90b}
{Soong}, Y., {Gruber}, D.~E., {Peterson}, L.~E., \& {Rothschild}, R.~E.
  1990{\natexlab{a}}, ApJ, 348, 641

\bibitem[{{Soong} {et~al.}(1990{\natexlab{b}}){Soong}, {Gruber}, {Peterson}, \&
  {Rothschild}}]{Soong_etal90}
{Soong}, Y., {Gruber}, D.~E., {Peterson}, L.~E., \& {Rothschild}, R.~E.
  1990{\natexlab{b}}, ApJ, 348, 634

\bibitem[{{Staubert} {et~al.}(2007{\natexlab{a}}){Staubert}, {Klochkov}, \&
  {Rodina}}]{Staubert_etal07b}
{Staubert}, R., {Klochkov}, D., \& {Rodina}, L. 2007{\natexlab{a}}, A\&A, [in
  preparation]

\bibitem[{{Staubert} {et~al.}(2006){Staubert}, {Schandl}, {Klochkov}, {Wilms},
  {Postnov}, \& {Shakura}}]{Staubert_etal06a}
{Staubert}, R., {Schandl}, S., {Klochkov}, D., {et~al.} 2006, in American
  Institute of Physics Conference Series, Vol. 840, The Transient Milky Way: A
  Perspective for MIRAX, ed. J.~{Braga}, F.~{D'Amico}, \& R.~E. {Rothschild},
  65--70

\bibitem[{{Staubert} {et~al.}(2007{\natexlab{b}}){Staubert}, {Shakura},
  {Postnov}, {Wilms}, {Rothschild}, {Coburn}, {Rodina}, \&
  {Klochkov}}]{Staubert_etal07}
{Staubert}, R., {Shakura}, N.~I., {Postnov}, K., {et~al.} 2007{\natexlab{b}},
  A\&A, 465, L25

\bibitem[{{Sunyaev} {et~al.}(1988){Sunyaev}, {Gilfanov}, {Churazov},
  {Loznikov}, {Efremov}, {Kaniovskii}, {Kuznetsov}, {Melioranskii}, {Voges},
  {Pietsch}, {Dobereiner}, {Engelhauser}, {Reppin}, {Trumper}, {Ogelman},
  {Kendizorra}, {Mony}, {Maisack}, {Staubert}, {Smith}, \&
  {Parmar}}]{Sunyaev_etal88}
{Sunyaev}, R.~A., {Gilfanov}, M.~R., {Churazov}, E.~M., {et~al.} 1988, Soviet
  Astronomy Letters, 14, 416

\bibitem[{{Tanaka}(1986)}]{Tanaka86}
{Tanaka}, Y. 1986, in Lecture Notes in Physics, Berlin Springer Verlag, Vol.
  255, IAU Colloq. 89: Radiation Hydrodynamics in Stars and Compact Objects,
  ed. D.~{Mihalas} \& K.-H.~A. {Winkler}, 198

\bibitem[{{Tananbaum} {et~al.}(1972){Tananbaum}, {Gursky}, {Kellogg},
  {Levinson}, {Schreier}, \& {Giacconi}}]{Tananbaum_etal72}
{Tananbaum}, H., {Gursky}, H., {Kellogg}, E.~M., {et~al.} 1972, ApJ, 174, L143

\bibitem[{{Tr\"umper} {et~al.}(1986){Tr\"umper}, {Kahabka}, {Oegelman},
  {Pietsch}, \& {Voges}}]{Truemper_etal86}
{Tr\"umper}, J., {Kahabka}, P., {Oegelman}, H., {Pietsch}, W., \& {Voges}, W.
  1986, ApJ, 300, L63

\bibitem[{{Tsygankov} {et~al.}(2006){Tsygankov}, {Lutovinov}, {Churazov}, \&
  {Sunyaev}}]{Tsygankov_etal06}
{Tsygankov}, S.~S., {Lutovinov}, A.~A., {Churazov}, E.~M., \& {Sunyaev}, R.~A.
  2006, MNRAS, 371, 19

\bibitem[{{Tsygankov} {et~al.}(2007){Tsygankov}, {Lutovinov}, {Churazov}, \&
  {Sunyaev}}]{Tsygankov_etal07}
{Tsygankov}, S.~S., {Lutovinov}, A.~A., {Churazov}, E.~M., \& {Sunyaev}, R.~A.
  2007, Astronomy Letters, 33, 368

\bibitem[{{Ubertini} {et~al.}(2003){Ubertini}, {Lebrun}, {Di Cocco}, {Bazzano},
  {Bird}, {Broenstad}, {Goldwurm}, {La Rosa}, {Labanti}, {Laurent}, {Mirabel},
  {Quadrini}, {Ramsey}, {Reglero}, {Sabau}, {Sacco}, {Staubert}, {Vigroux},
  {Weisskopf}, \& {Zdziarski}}]{Ubertini_etal03}
{Ubertini}, P., {Lebrun}, F., {Di Cocco}, G., {et~al.} 2003, A\&A, 411, L131

\bibitem[{{Vedrenne} {et~al.}(2003){Vedrenne}, {Roques}, {Sch{\"o}nfelder},
  {Mandrou}, {Lichti}, {von Kienlin}, {Cordier}, {Schanne}, {Kn{\"o}dlseder},
  {Skinner}, {Jean}, {Sanchez}, {Caraveo}, {Teegarden}, {von Ballmoos},
  {Bouchet}, {Paul}, {Matteson}, {Boggs}, {Wunderer}, {Leleux},
  {Weidenspointner}, {Durouchoux}, {Diehl}, {Strong}, {Cass{\'e}}, {Clair}, \&
  {Andr{\'e}}}]{Vedrenne_etal03}
{Vedrenne}, G., {Roques}, J.-P., {Sch{\"o}nfelder}, V., {et~al.} 2003, A\&A,
  411, L63

\bibitem[{{Voges}(1985)}]{Voges84}
{Voges}, W. 1985, NASA STI/Recon Technical Report N, 85, 34112

\bibitem[{{Voges} {et~al.}(1982){Voges}, {Pietsch}, {Reppin}, {Tr\"umper},
  {Kendziorra}, \& {Staubert}}]{Voges_etal82}
{Voges}, W., {Pietsch}, W., {Reppin}, C., {et~al.} 1982, ApJ, 263, 803

\bibitem[{{Vrtilek} \& {Halpern}(1985)}]{VrtilekHalpern85}
{Vrtilek}, S.~D. \& {Halpern}, J.~P. 1985, ApJ, 296, 606

\bibitem[{Wilms {et~al.}(2003)Wilms, Ketsaris, Postnov,
  {et~al.}}]{Wilms_etal03}
Wilms, J., Ketsaris, N.~A., Postnov, K.~A., {et~al.} 2003, Izvestiya Akademii
  Nauk, Ser. Fizicheskaya, 67, 310

\bibitem[{{Winkler} {et~al.}(2003){Winkler}, {Courvoisier}, {Di Cocco},
  {Gehrels}, {Gim{\'e}nez}, {Grebenev}, {Hermsen}, {Mas-Hesse}, {Lebrun},
  {Lund}, {Palumbo}, {Paul}, {Roques}, {Schnopper}, {Sch{\"o}nfelder},
  {Sunyaev}, {Teegarden}, {Ubertini}, {Vedrenne}, \& {Dean}}]{Winkler_etal03}
{Winkler}, C., {Courvoisier}, T.~J.-L., {Di Cocco}, G., {et~al.} 2003, A\&A,
  411, L1

\end{thebibliography}

\end{document}